\documentclass[a4paper,12pt]{article}
\usepackage{jcappub}
\usepackage{amsthm,graphicx,multirow}
\usepackage{epsfig}
\usepackage{latexsym, amssymb} 
\usepackage{amsmath}
\def\beq{\begin{equation}}
\def\eeq{\end{equation}}
\def\br{\begin{eqnarray}}
\def\er{\end{eqnarray}}
\def\benu{\begin{enumerate}}
\def\efnu{\end{enumerate}}
\def\nn{\nonumber}

\def\l{\left}
\def\r{\right}

\def\d{{\rm d}}

\def\vka{{\bf k}_{1}}
\def\vkb{{\bf k}_{2}}
\def\vkc{{\bf k}_{3}}

\def\cB{{\cal B}}
\def\fnl{f_{_{\rm NL}}}

\def\cl{{\cal C}_{\ell}}

\begin{document}
\begin{center}
\title{Wiggly Whipped Inflation } 
\end{center}
\author[a]{Dhiraj Kumar Hazra,} 
\author[a,b]{Arman Shafieloo,}
\author[c,d]{George F. Smoot,}
 \author[e,f]{Alexei A. Starobinsky}

\affiliation[a]{Asia Pacific Center for Theoretical Physics, Pohang, Gyeongbuk 790-784, Korea}
\affiliation[b]{Department of Physics, POSTECH, Pohang, Gyeongbuk 790-784, Korea}
\affiliation[c]{Paris Centre for Cosmological Physics, APC (CNRS), Universit\' e Paris Diderot, \\
  Universit\'e Sorbonne Paris Cit\'e,  75013 France}
\affiliation[d]{Physics Department and Lawrence Berkeley National Laboratory, University of California, Berkeley, CA 94720, USA}
\affiliation[e]{Landau Institute for Theoretical Physics RAS, Moscow, 119334, Russian Federation}
\affiliation[f]{Kazan Federal University, Kazan 420008, Republic of Tatarstan, Russian Federation}
\emailAdd{dhiraj@apctp.org, arman@apctp.org, gfsmoot@lbl.gov, alstar@landau.ac.ru} 

\abstract 
{Motivated by BICEP2 results on the CMB polarization
B-mode which imply primordial gravitational waves are produced
when the Universe has the expansion rate of about $H \approx
10^{14}$ GeV, and by deviations from a smooth power-law behavior
for multipoles $\ell <50$ in the CMB temperature anisotropy power
spectrum found in the WMAP and Planck experiments, we have
expanded our class of large field inflationary models that fit
both the BICEP2 and Planck CMB observations consistently. These
best-fitted large field models are found to have a transition from
a faster roll to the slow roll $V(\phi)=m^2 \phi^2/2$ inflation at
a field value around 14.6~${\rm M_{Pl}}$ and thus a potential
energy of $V(\phi) \sim (10^{16}\,{\rm GeV})^4$. In general this
transition with sharp features in the inflaton potential produces
not only suppression of scalars relative to tensor modes at small
$k$ but also introduces wiggles in the primordial perturbation
spectrum. These wiggles are shown to be useful to explain some
localized features in the CMB angular power spectrum and can also
have other observational consequences. Thus, primordial GW can be
used now to make a tomography of inflation determining its fine
structure. The resulting Wiggly Whipped Inflation scenario is described in
details and the anticipated perturbation power spectra, CMB power
spectra, non-Gaussianity and other observational consequences are
calculated and compared to existing and forthcoming observations.}

\maketitle
\section{Introduction}
The recent BICEP2 report~\cite{BICEP2:Detection,BICEP2:datasets}
of a CMB polarization B-mode signal consistent with the signature
of primordial gravitation waves (GW), when combined with previous
CMB temperature anisotropy data~\cite{Planck:lilelihood}, has two
consequences for the inflationary scenario of the early Universe:
(1) these GW are produced at the Universe expansion rate
$H(t)\equiv \dot a(t)/a(t)$ of about $10^{14}$ GeV in energy units
and (2) the inflationary model has to be modified to the extent of
adding more parameters beyond the only one required for its
simplest realizations. As a result, the concordance model of the
contemporary Universe acquires more parameters, too. These are
important consequences and their strength depends upon the
confirmation and improvement of the BICEP2 results. We proceed
under the assumption that the BICEP2 report is essentially correct
for the purposes of this paper.

In the context of the inflationary mechanism of primordial
gravitational GW production, what ultimately follows from the
BICEP2 measurement is the expansion rate $H(t(k))$ for the range
of comoving wave vectors $k$ corresponding to the multipoles $\ell
= 30-100$ at the time around their first Hubble radius crossing,
$k=aH$, during the early quasi-de Sitter (inflationary) stage.
Tensor perturbations (to be primordial GW after the second Hubble
radius crossing much later, at the radiation or even recent matter
dominated stages) arise from quantum vacuum fluctuations of the
gravitational field at this stage. Their amplitude is determined
by $H$ during inflation, or it can be said, by the de Sitter
(Gibbons-Hawking) temperature $T=H/2\pi$, though the energy
spectrum of primordial GW after the second Hubble radius crossing
is strongly non-thermal that makes possible their detection at
cosmological scales very much exceeding the thermal scale of CMB.
The power spectrum of primordial metric tensor perturbations
generated during the quasi-de Sitter stage, first calculated
in~\cite{Starobinsky:1979} where the final answer was presented in
the equivalent form of the spectral energy density of GW after the
second Hubble radius crossing, is given by \beq P_T(k)\equiv
rP_S(k)=\frac {2H(k)^2}{\pi^2 M_{Pl}^2} \eeq where $H(k)$ is the
expansion rate $H(t)$ estimated at the moment when $k=aH$ during
inflation and $M_{\rm Pl}$ is the reduced Planck mass, 
$M_{\rm Pl} = \sqrt{\hbar c / 8 \pi G} = 2.435 \times 10^{18} {\rm GeV}/c^2$.
Using the normalization $P_S = 2.2 \times 10^{-9}$ at the pivot
scale $k =$0.05 Mpc$^{-1}$ (note that GW contribution to CMB
anisotropy is negligible for multipoles corresponding to this
scale), more precisely we have \beq H(k_{\ast}) = 5.0 \times
10^{-5} \left(\frac {r_{0.002}} {0.2}\right)^{1/2} 5^{(0.96-n_{\rm
S})} M_{Pl} \approx 10^{14}\,{\rm GeV}\eeq where $r_{0.002}$ is
the tensor-to-scalar ratio at $k_{\ast} = 0.002$ Mpc$^{-1}$ and
$n_{\rm S}$ is the scalar perturbation spectral index.
$H(k_{\ast})$ determines the characteristic energy scale of
inflaton scalar particles and gravitons at the time of their
creation. It is much less than the energy density scale $\sim
10^{16}$ Gev (the GUT scale) of the inflaton potential that
reflects the fact that inflation is ``cold".

Moreover, the inflaton mass during slow-roll inflation should be
in turn significantly less than $H$. In particular, would we
restrict ourselves to the $\ell >50$ CMB anisotropy data from
Planck~\cite{Planck:lilelihood} and the CMB polarization B-mode
data from BICEP2~\cite{BICEP2:Detection,BICEP2:datasets}, then the
simplest inflaton model with $V(\phi)=m^2\phi^2/2$ and the
constant inflaton mass $m\approx 2\times 10^{13}$ GeV would
produce a very good fit to these data for the standard number of
light neutrino species, $N_{\nu}=3$ (in this case, $H\approx
m\phi/\sqrt{6}M_{Pl}$). So, even in this simplified approach
inflation requires inflaton masses much less than the GUT scale.

Now, with the three CMB anisotropy spectra, temperature and E and
B-mode polarizations (actually TT, TE, EE, and BB power spectra),
we can go much further and determine the inflaton potential
$V(\phi)$, its slope and, even more important for particle
physicists, the effective inflaton mass, $m_{eff}^2=V''(\phi)$,
irrespective of the knowledge of an underlying microscopic field
(string, M-, etc.) theory. In particular, it will be shown in our
paper below that taking into account the features in the CMB
anisotropy spectrum observed for multipoles $\ell\lesssim 40$
which are of the {\em same order}, $\sim 10\%$, as the relative
contribution of primordial GW with $r\sim 0.2$ to CMB anisotropy,
leads to a whole range of inflaton effective masses from $2\times
10^{13}$ GeV to $10^{14}$ GeV and even more.

The concordance model of cosmology is based on the assumption of
the power-law form of the primordial scalar perturbation spectrum
and the spatially flat $\Lambda$CDM background FLRW model. Though
there were hints of deviation from the concordance model since
WMAP first year data~\cite{Peiris:2003ff}, this model was
consistent with the data within uncertainties of observations.
With Planck it has been shown that the data indicate significant
deviation from the concordance model at small $k$ ($\ell < 50$)
~\cite{Hazra:concordance}. At the same time the data indicate
certain localized features in the CMB angular spectrum  both for
$\ell =2,3$ and in the range $20\lesssim \ell \lesssim 40$.
However, the presence of the large scale features could not be
confirmed by CMB temperature data alone due to cosmic variance. On
the other hand, the BICEP2
data~\cite{BICEP2:Detection,BICEP2:datasets}, when combined with
the Planck temperature data~\cite{Planck:lilelihood}, indirectly
confirm these large scale features, most importantly a strong
suppression in the scalar primordial power spectrum (PPS) in the range
$20\lesssim \ell \lesssim 40$ at more than
3$\sigma$~\cite{Hazra:recent}, since in the presence of primordial
GW the required suppression in the scalar spectrum becomes {\em
larger}. This confirmation certainly opens up a possibility to
look for particular inflationary models that provides this
suppression~\cite{Hazra:recent,Hazra:2014jka,Contaldi:2014zua,
Miranda:2014wga,Abazajian:2014tqa,Hu:2014aua}.
In~\cite{Hazra:2014jka} we have discussed canonical inflationary
models, Whipped Inflation, that can generate the large scale
scalar suppression. Afterwards, different inflationary models have
been discussed in the context of reconciling Planck and
BICEP2~\cite{Kawasaki:2014fwa,Freivogel:2014hca,Zibin:2014iea,
Achucarro:2014msa,Bousso:2014jca,Firouzjahi:2014fda,Kinney:2014jya,
Kim:2014rwa,Kallosh:2014xwa,Mukohyama:2014gba}.

Though there have been efforts to reconcile the observations using
an additional neutrino~\cite{Dvorkin:2014lea} and non-Bunch-Davies
vacuum~\cite{Ashoorioon:2014nta} (which effectively changes the
low-$\ell$ power law), we remain firmly convinced that modifying
the power law for the low $k$ scalar perturbation spectrum is
probably the most likely and reasonable approach which has been
discussed in \cite{Hazra:recent,Hazra:2014jka}.


In this paper, we go a step beyond our work
in~\cite{Hazra:2014jka} and naturally extend the scope of the
Whipped Inflation potential. Implementing sharp features of the
inflaton potential like its (smoothed) step-like behavior or a
rapid change of its first derivative into Whipped Inflation, we
show that the new models generates oscillations/wiggles along with
suppression in the primordial scalar power spectrum at large
scales (small $k$). These oscillations along with the suppression
fits the Planck angular power spectrum of temperature anisotropy
both at low $\ell < 50$ and high $\ell$ better than the
concordance model of cosmology. The wiggles in the scalar 
primordial power spectra are imprinted on the matter power 
spectrum too, which would change the
large scale structure observables. From our analysis we identify
two models and we estimate the probabilities to detect the
features in the scalar primordial power spectrum using large scale
structure data from a survey such as Dark Energy Spectroscopic Instrument 
(DESI)~\cite{Levi:2013gra,DESI}~\footnote{DESI is descended from BigBOSS and 
aimed at obtaining the optical spectra of galaxies and quasars}.

The paper is organized as follows. In Sec.~\ref{sec:tension} we
describe the tension between Planck temperature anisotropy
observation and BICEP2 B-mode observation within the context of
power law form of the scalar primordial power spectrum and mention
the possible ways to reconcile them. In Sec.~\ref{sec:scenario} we
provide the Wiggly Whipped Inflation scenarios, and construct the
potential we use in this work. Sec.~\ref{sec:num} briefly
discusses the essential numerical details used to solve the
potential and to compare the scalar and the tensor PPS to the
data. Sec.~\ref{sec:results} provides the results of our analysis,
confronting the proposed theoretical models to the Planck, BICEP2
and other datasets and also discussing the non-Gaussianities
generated during inflation. We do also a forecast analysis,
deriving the shape of the matter power spectrum for the proposed
models comparing them with the expectations of the concordance
model and see how well we can distinguish these models from each
other using the sensitivity of the future DESI experiment. In
Sec.~\ref{sec:conclusions} we summarize.

\section{CMB Temperature and B-mode Polarization Tension}~\label{sec:tension}

The Planck observed temperature anisotropy power spectrum is in tension with the BICEP2 B-mode polarization spectrum in the context of the 
concordance model of $\Lambda$CDM with a power-law scalar perturbation spectrum. 
Planck low-$\ell$ TT data was previously $\sim$10\% lower than the best-fitted model and, 
if the B-mode polarization is interpreted as tensor modes, 
then the tensor modes should add low-$\ell$ temperature anisotropy power. 
Due to the observed suppressed low-$\ell$ TT power spectrum, 
Planck indicates low tensor-to-scalar ratio ($r<0.11$ at 95\% CL)~\cite{Ade:2013uln}, 
while BICEP2 indicates much higher $r$ ($\sim 0.2$). 
A reasonable way to  address the tension is to bring in features in the scalar primordial power spectrum.
This in turn reflects that the single power law form of the primordial spectrum is not supported by the observations. 
Hence, a modification of the power law scalar PPS becomes necessary.

\subsection{Necessity to Modify the Power Law Form}

Scalar primordial power spectra can be modified by keeping an eye to the inconsistencies in the datasets 
that are not addressed by conventional models. 
Parametrization of the primordial power spectra and model independent 
reconstruction~\cite{Hazra:2013nca,reconstruction-all} can reveal the position of the features indicated by the data. 
We list a few possibilities to address the feature in the scalar PPS from phenomenological and theoretical point of view.

\subsubsection{Running the power law spectral index} 
The power law form of the primordial spectrum, $P_{\rm S}(k)$, is described by, 
\beq
P_{\rm S}(k) = A_{\rm S} (k / k_{\ast})^{n_{\rm S}-1},~\label{eq:plaw-pps}
\eeq
where, $A_{\rm S}$ is the amplitude at the pivot scale $k_{\ast}$ and $n_{\rm S}$ is the tilt of the spectrum.
Planck TT data constrains the value $n_{\rm S}\sim 0.9603\pm0.0073$. One can consider allowing the index to vary with $k$ so that 
$n_{\rm S}(k) = n_{\rm S}(k_{\ast}) + d n_{\rm S}(k)/ d \ln~k $. The running scalar spectral index have been  
used in order to reconcile the Planck and BICEP2 data~\cite{BICEP2:Detection,Abazajian:2014tqa}. We need 
sufficiently large running $d n_{\rm S} / d \ln~k \sim -0.02$ in order to match the data which in turn modifies 
the small scale power as well. To calibrate the small scale power, we then need another degree of freedom (neutrino 
mass, running of running, etc.). Running provides an improvement of $-2\Delta\ln{\cal L}\sim-6.5$ and indicates that 
power law scalar PPS is rejected by more than $2\sigma$. 
However,  the data seem just to require
 a PPS with a suppression in the large scales only, which is achievable by a break in the power law. 

\subsubsection{Broken power law}
Introducing a break in the power spectrum (two different slopes or powers) shows that 
with one extra parameter the broken  the PPS is supported by the data at more than 3$\sigma$ CL, 
compared to power law PPS~\cite{Hazra:recent}. 
The broken PPS can provide $-2\Delta\ln{\cal L}\sim -12.5$ improvement in fit
compared to power law. 
The extra one degree of freedom makes the PPS substantially more flexible compared to running spectral index without altering small scale power.
Theoretically different behavior of scalar field in early and late stages of inflation can describe this phenomenological PPS.
With the help of Whipped Inflation we have shown that similar power spectrum is indeed achievable with 
low level of non-Gaussianities and large tensors~\cite{Hazra:2014jka}. 

\subsubsection{Step in the primordial perturbation spectrum}

We~\cite{Hazra:recent} and others~\cite{Contaldi:2014zua,Miranda:2014wga} also showed that a step in the primordial 
perturbation scalar spectrum would also provide a good fit to the Planck TT power spectrum. 
We show here a displaced potential at a critical scale can easily produce such a step in the PPS
and the results fit the Planck and BICEP2 observations better than a simple fixed power law.
If the transition to the displacement is sharp, then there are wiggles introduced to the higher-$k$ modes 
but these still fit the observations well.

\subsubsection{Step in the Inflaton potential}
A step in the inflation potential where the power law breaks can generate an intermediate 
fast-roll phase which can produce localized wiggles in the scalar PPS and the TT angular power spectra~\cite{step-models}. 
This helps us to fit a few features in the angular power spectra near 
$\ell=22$ and 40. In this paper we shall demonstrate that a simple extension of Whipped Inflation can naturally 
address the large scale suppression that can reconcile Planck and BICEP2 data along with the generation of wiggles in the scalar PPS. 
The wiggles obtained from the extension can address the localized features in the TT data. 
This has now become important since from BICEP2, any deviation at large scale power is now more favored than the power law. Moreover,
the future polarization data and large scale structure data, with their projected sensitivity, can certainly falsify the existence of 
these features.

\subsection{Getting to the Inflation Potential}
Here we outline what we expect from the Inflaton potential in order to satisfy different observables. 
\begin{enumerate}
 \item Suppression of low-$\ell$ scalars but with large amplitude tensor perturbations.
 
 \item A model that puts a feature on a scale between the horizon to (1/100)'th of the horizon. 
 
  \item Resume complete slow roll around $\ell > 100$ and have it persist for an extended ($N> 50$) e-folds.
  
  \item Generate negligible non-Gaussianities~\cite{Planck:fnl} or non-Gaussianities that would have been over-looked until now.
\end{enumerate}

To begin with, let us consider the suppression of the low-$\ell$ scalars but with large tensor perturbations.
The scalar power spectrum from Inflation is given by $P_{\rm S}(k) = A_{\rm S} k^{n_{\rm S}-1}$ and the amplitude 
$P_{\rm S}(k) \propto V(\phi)^3 / V_{\phi}(\phi)^2$ while for the tensor perturbations $P_T(k) = A_{\rm T} k^{n_{\rm T}}$ where the amplitude 
$P_{\rm T} \propto H^2 \propto V(\phi)$ where $V(\phi)$ is the inflation potential. To have significant, a la BICEP2, tensors one 
would need to have a relatively steep potential and thus a large field inflation. However pushing $V$ up would make the 
scalars go up significantly unless one makes $V_\phi(\phi) \equiv dV/d\phi$ increase by enough more to over compensate
and reduce the scalars by the 10 to 15 per cent needed. That takes one from the slow roll regime 
into the fast roll regime and then one needs to transit to the slow roll regime to get sufficient 
{\it e-folds} to generate the full primordial perturbation spectrum and to take care of other issues.

That approach takes care of both points one and two. In order to get to slow roll (point 3) one must then 
transition to a less steep potential. Hence the need to break the potential from a steep one to a significantly less steep one
at a preferred scale to make the transition from the low-$\ell \sim 100$ or low $k$ to the higher $\ell$ or $k$ portion.

We were concerned that such an abrupt transition would generate significant non-Gaussianity, which is constrained under
Planck observations. We were able to make such a model and avoid the significant ringing that generically accompany abrupt features.
However, we have since realized that some ringing ``wiggles" can actually describe the data more precisely than the more damped versions.

\section{Wiggly Whipped Inflationary Scenario}~\label{sec:scenario}
In the paper~\cite{Hazra:2014jka} we had introduced Whipped Inflation potential. 
In this paper, motivated by possible Whipped Inflation scenarios and keeping an eye to the 
sharp features in the temperature anisotropy data, 
we introduce a first order and second order transition in the Whipped
inflation potential, which we call Wiggly Whipped Inflaton potential. For similar types of transitions 
that were discussed in literature, see~\cite{S92,Linde:1998iw,Linde:1999wv,JSS08,JSSS09,Bousso:2013uia}.

\subsection{Wiggly Whipped first-Order Transition}

In this transition we introduce a jump in the potential of the inflaton field {\it at} the transition ($\phi_0$), 
given by Eq.~\ref{eq:potential_o1}. 

\begin{equation}
V({\phi})=\gamma\phi^p+\lambda\left[(\phi-\phi_0)^q+\phi_{01}^q\right]~\Theta(\phi-\phi_0),~\label{eq:potential_o1}
\end{equation}
Note that for $\phi_{01}=0$, the potential simply reduces to the 
Whipped Inflaton potential that we proposed in the recent paper~\cite{Hazra:2014jka}. Since the field starts rolling from 
a steeper power law potential and smoothly transits to a flat power law potential, we find a mild departure from initial 
slow-roll phase, imprinting a large scale suppression in scalar primordial power spectra. 
The modified  Whipped Inflation potential with a discontinuity $\phi_{01}^q$ in the potential at $\phi_0$, ensuring 
a momentary intermediate boost in the kinetic energy of the scalar field during the inflation. 
The boost imprints wiggles/oscillations in the primordial power spectra. 
One can certainly expect significantly large non-Gaussianities as well from the model. Since a discontinuity in the potential
gives rise to divergent derivative in the potential at the transition, we smooth the discontinuity. We have smoothed the discontinuity
using step functions such as $1+{\rm erf}[{(\phi-\phi_0)}/{\Delta}]$ and $1+{\tanh}[{(\phi-\phi_0)}/{\Delta}]$. The 
features in the primordial scalar power spectrum and the bispectrum depend strongly on the width of the transition $\Delta$. In this paper 
we have only considered $(p,q)=(2,3)$ since for Whipped Inflation we obtained best fit in this combination~\cite{Hazra:2014jka}.

\subsection{Wiggly Whipped second-Order Transition}

The second order transition in the inflaton potential originally appeared in~\cite{S92} for linear potential. Primordial 
features and non-Gaussianities generated by this inflaton potential have been discussed widely in the 
literature~\cite{Takamizu:2010,Martin:2011sn,Arroja:2011,Hazra:2012yn,Arroja:2012,Martin:2014}. Since BICEP2 data indicated a 
large tensor-to-scalar ratio, we revisit similar 
transition in the context of Whipped Inflation. The potential that we consider in this paper is given by~\ref{eq:potential_o2}.

\begin{equation}
V({\phi})=\gamma\phi^p+\lambda\phi^p~(\phi-\phi_0)~\Theta(\phi-\phi_0),~\label{eq:potential_o2}
\end{equation}
Note that, the potential here is continuous at $\phi_0$ but not its derivatives. Here also, following the first order transition  
we smooth the derivative of the potential with a step function of width $\Delta$. Here, we work with $p=2$, quadratic inflation, 
we are able to generate appropriate tensor amplitude that is supported by BICEP2 data. The scalar PPS generated in this model comes close 
to providing a step in the primordial scalar perturbation spectrum. 

\section{Essential numerical details}~\label{sec:num}

Background inflationary equations and the scalar and tensor perturbation equations for the Wiggly Whipped potential are calculated 
using the publicly available code  BI-spectra and Non-Gaussianity Operator, {\tt BINGO}~\cite{Hazra:2012yn}. 
Allowing sufficient {\it e-folds} $\sim60-70$ and by using initial slow roll we fixed the inflaton initial condition. 
Initial scale factor is estimated assuming that the pivot scale $k_{\ast}=0.05~{\rm Mpc^{-1}}$ leaves the Hubble radius 50 {\it e-folds} before the 
end of inflation. We have modified {\tt CAMB}~\cite{cambsite,Lewis:1999bs} to work with the {\tt BINGO} outputs directly. 
To find the best fit we have Powell's BOBYQA method of iterative minimization~\cite{powell} through {\tt COSMOMC}~\cite{cosmomcsite,Lewis:2002ah}. 
The {\tt commander} and {\tt CAMspec} likelihood are used to estimate the low-$\ell$
and high-$\ell$ likelihood from Planck data~\cite{Planck:lilelihood} respectively. We have used WMAP low-$\ell$ (2-23) E-mode
polarization data~\cite{Hinshaw:2012fq}
(denoted as WP in results section). The complete BICEP2 likelihood is calculated using bandpowers for 9 bins for E and
B mode polarization data. 
We should also mention here that to make our analysis robust, we have allowed the background cosmological parameters and the 14 Planck 
foreground nuisance parameters to vary along with the inflationary potential parameters. Note that for Planck, the estimation of 
the best fit should be done in few steps, since the large number of parameters (inflationary potential parameters + 4 background cosmological 
parameters + 14 nuisance parameters) often lead the method to a local minima in the parameter space. To get the best fit,
we have performed our search by changing 
the initial starting points in the parameter spaces. In this paper, we shall present two such minima obtained for the first order 
transition (Eq.~\ref{eq:potential_o1}) attempting to fit different features in the angular power spectra. 
Moreover, it should be noted that since by default {\tt camb} calculates the angular power spectra
in few multipoles and interpolate between them, to ensure that the wiggles in the scalar PPS are not missed by the interpolation, we perform 
our analysis by calculating the angular power spectra in every multipole. The matter power spectra that we present in our analysis for the best fit values of 
the parameters are also calculated using {\tt camb}. To calculate the non-Gaussianity for this inflationary model, we 
again use {\tt BINGO} in the equilateral limit. In all our analyses we have assumed spatially flat FLRW Universe. In all our analyses we have assumed 
spatially flat FLRW Universe. We have defined ${\rm M_{Pl}}^2=1/(8\pi G)$ and used $\hslash=c=1$ throughout the paper. 


\section{Results and discussions}\label{sec:results}

\subsection{Best fit results}

We start this section by tabulating the best fit results from Wiggly Whipped potential, 
corresponding to Eq.~\ref{eq:potential_o1} and~\ref{eq:potential_o2} in 
Table~\ref{tab:bestfits}. For the first order transition, we provide 2 results corresponding to two 
minima in the parameter spaces. We note that the width of the step in 
the first order transition affects the scalar primordial power spectrum severely. Our searches in the
parameter space revealed that a relatively smooth step (First order - I) 
leads to an improvement in the low-$\ell$ TT angular spectrum from Planck and a sharp transition (First order - II) 
attempts to fit the high-$\ell$ glitches in the TT power spectrum, unaddressed by the power law form of scalar PPS.
In fact it had been demonstrated in some earlier works~\cite{oip} that violent oscillations in the scalar primordial power spectrum might help to fit the CMB data better than power law model.

\renewcommand{\arraystretch}{1.1}
\begin{table*}[!htb]
\begin{center}
\vspace{4pt}
\begin{tabular}{|c | c | c | c |}
\hline\hline
 \multicolumn{4}{|c|}{{\bf Best fit inflation potential (Eqs.~\ref{eq:potential_o1} and~\ref{eq:potential_o2}) and cosmological parameters}}\\

\hline
& First order - I & First order - II & Second Order\\

\hline

$\Omega_{\rm b}h^2$ & 0.022&0.0219 & 0.0219\\
\hline

$\Omega_{\rm CDM}h^2$ & 0.1203& 0.1213& 0.1205\\
\hline

$100\theta$ &  1.041& 1.04 & 1.041\\
\hline

$\tau$ & 0.097& 0.085& 0.1\\
\hline

$\gamma$& $2.65\times10^{-11}$ & $2.59\times10^{-11}$&$2.68\times10^{-11}$\\
\hline

$\lambda$ & $2.13\times10^{-10}$ &$3.63\times10^{-10}$& $5.2\times10^{-13}$\\
\hline

$\phi_0$ in $\rm M_{\rm Pl}$ & $14.66$ & 14.69 & 14.59\\
\hline
$\phi_{01}$ in $\rm M_{\rm Pl}$ & $0.52$ & 0.18 & -\\

\hline

\hline\hline

$\Omega_{\rm m}$ & 0.32&0.33  & 0.32\\
\hline
$H_{0}$ & 66.9& 66.4& 66.8\\
\hline\hline
\multicolumn{4}{c}{$-2\ln{\cal L}$ [Best fit]}\\
\hline
{\tt commander} [-1.13] & -13.42 &-1.44  &-9.67\\
Planck ($\ell=2-49$) &  &  &\\
\hline
{\tt CAMspec} [7797.29]& 7795.68 & 7789.24& 7794\\
Planck ($\ell=50-2500$) &  & & \\

\hline
{\tt WP} [2013.38]& 2014.34 & 2013.39& 2014.1\\
\hline
{\tt BICEP2} [40.04]& 39.56 & 39.8& 39.4\\
\hline
Total [9849.58]& 9836.16 & 9841 &9837.8\\
\hline
$-2\Delta\ln{\cal L}$ & -13.42 &-8.59  & -11.8\\
\hline

\hline\hline
\end{tabular}
\end{center}
\caption{~\label{tab:bestfits} Best fit parameters for the Wiggly Whipped Inflaton potential 
Eqs.~\ref{eq:potential_o1} and~\ref{eq:potential_o2}.   
and the best fit cosmological parameters when compared with Planck + WP + BICEP2 data
combination. The improvement in fit, $-2\Delta\ln{\cal L}$ is 
obtained upon comparing the $\chi^2$ of the Wiggly Whipped scenario with the power law scalar PPS. The quantities in the square 
brackets in the likelihood section denotes the best fit likelihood for power law PPS, mentioned in~\cite{Hazra:recent}.} 
\end{table*}

Note that First order - I fits the CMB data from Planck + WP + BICEP2 significantly better ($-2\Delta\ln{\cal L}\sim -13.5$) 
than the power law PPS (for the best fit power law,
see~\cite{Hazra:recent}). Compared to Whipped Inflation, Wiggly Whipped performs better since the wiggles at the large scale 
scalar PPS fits the features around $\ell=22$ and $40$.  First order - II, on the other hand 
contains a sharp step in the potential which leads to violent oscillations in the primordial and the angular power spectrum. 
We find that the minima obtained 
around the sharp step region does not help to fit the low-$\ell$ data better and in that sense is not particularly interesting 
to reduce the tension between Planck
and BICEP2. However, compared to power law scalar PPS, this wiggly scalar PPS fits the high-$\ell$ {\tt CAMspec} likelihood 
better and provides 
an overall improvement of $-2\Delta\ln{\cal L}\sim -8.6$. We should mention that in the first order transition there exist 
a large degeneracy 
between the steepness of the potential at the initial stages of inflation ($\lambda$) and the extent of the discontinuity
in the potential $\phi_{01}^p$,
which is reflected in the table. However, to explore the complete degeneracy we need to have a full Markov Chain Monte 
Carlo analysis for the models which is beyond the scope of this paper.

The second order Wiggly Whipped potential provides an overall improvement 12 compared to power law PPS. Interestingly, 
the improvement comes both from low-$\ell$ and high-$\ell$. Hence, we get improvement both from {\tt commander} and {\tt CAMspec} 
which is certainly interesting.  

\subsection{Primordial scalar and tensor perturbation power spectrum}

\begin{figure*}[!htb]
\begin{center} 
\resizebox{210pt}{160pt}{\includegraphics{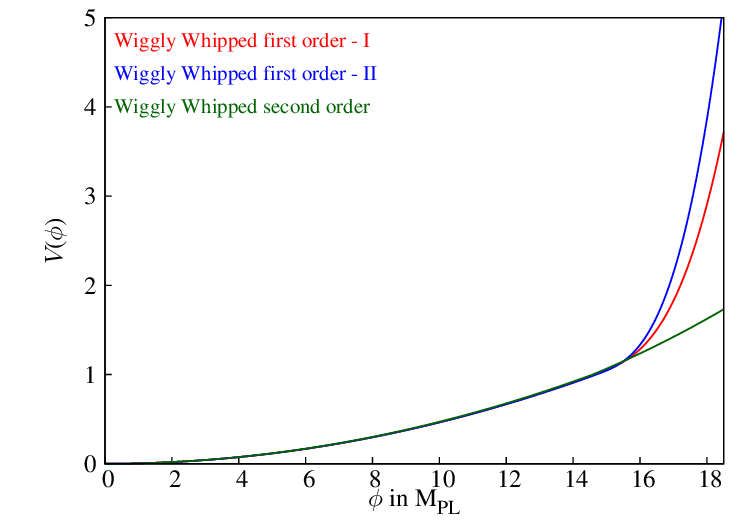}} 
\resizebox{210pt}{160pt}{\includegraphics{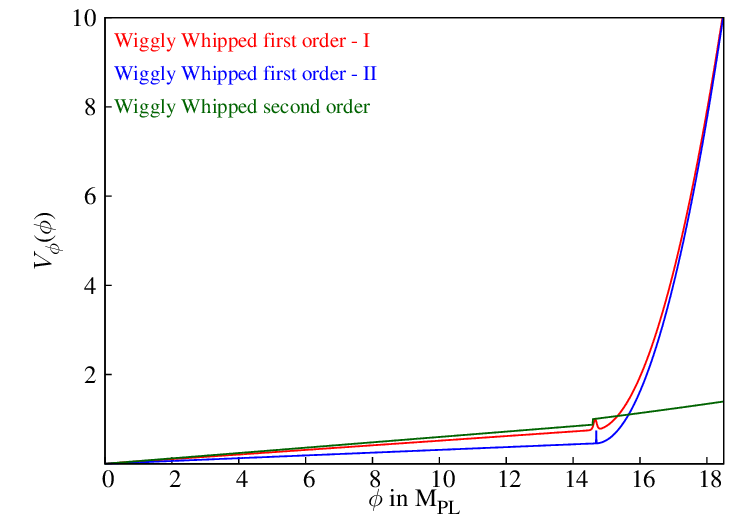}} 
\resizebox{210pt}{160pt}{\includegraphics{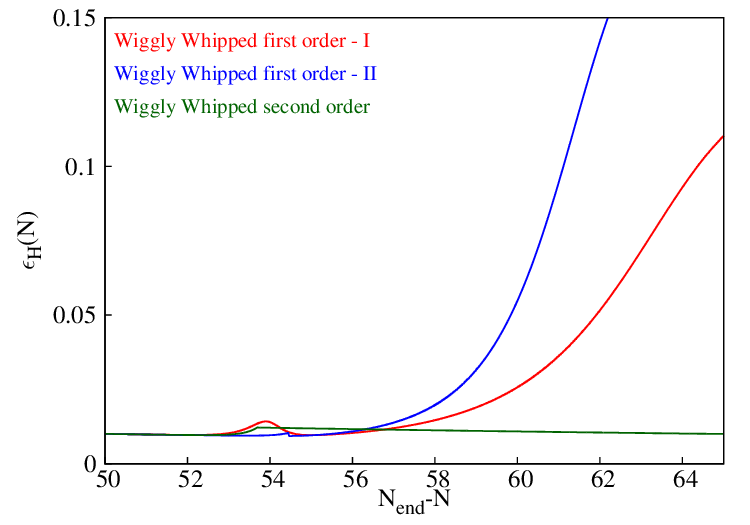}} 
\resizebox{210pt}{160pt}{\includegraphics{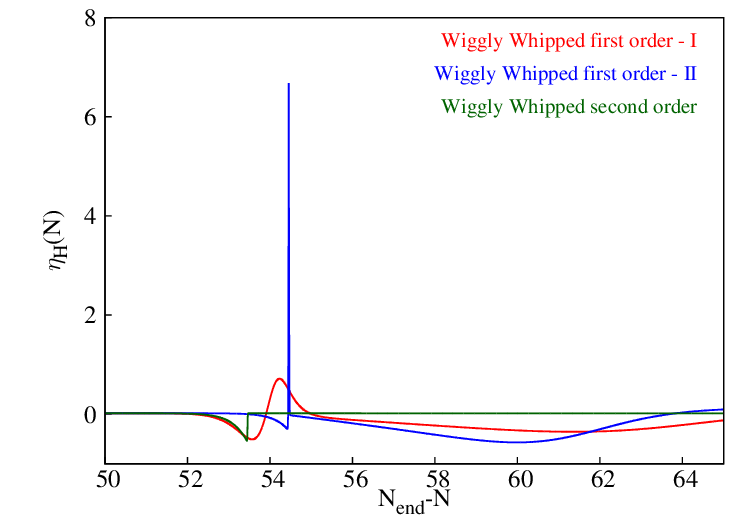}} 
\resizebox{210pt}{160pt}{\includegraphics{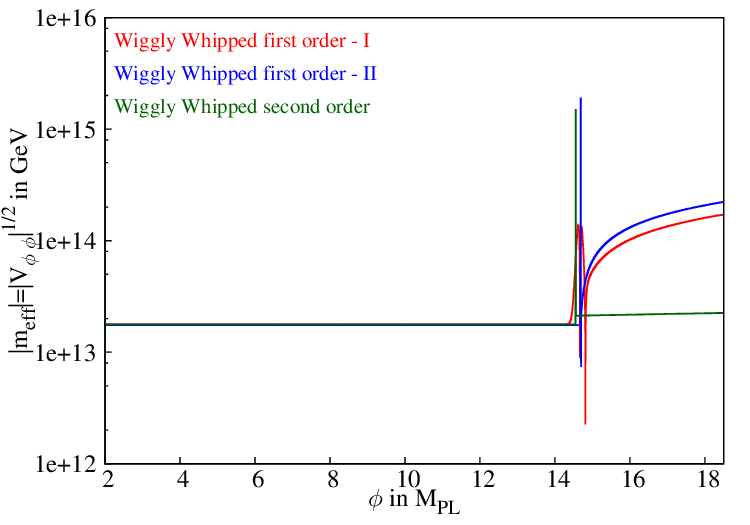}}
\resizebox{210pt}{160pt}{\includegraphics{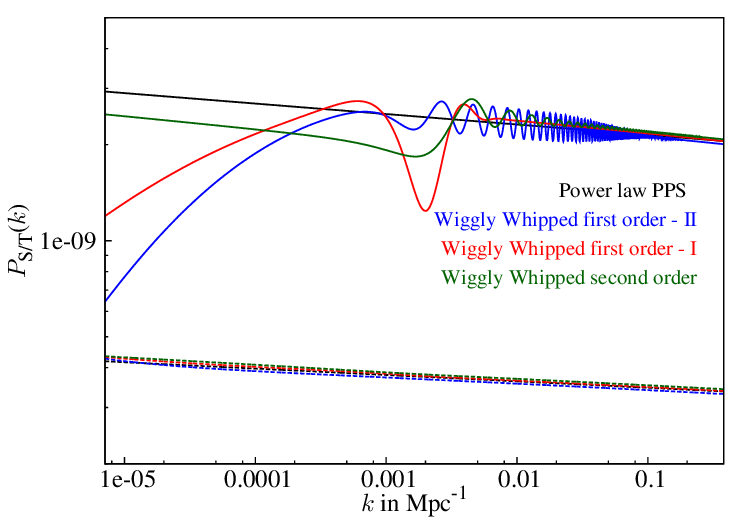}} 
\end{center}
\caption{\footnotesize\label{fig:theory} Wiggly Whipped Inflation : [Top]- The best fit potentials (left) and their derivatives (right) 
corresponding to Eqs.~\ref{eq:potential_o1} and~\ref{eq:potential_o2}. [Middle]- The best fit first slow roll parameter (left) and 
second slow roll parameter (right). [Bottom] The absolute value of effective mass of inflaton (left) and 
the primordial scalar (solid) and tensors (dashed) power spectra (right).}
\end{figure*}

In Figure.~\ref{fig:theory} we plot the relevant quantities for the best fit values denoted in Table~\ref{tab:bestfits}. At the top left panel 
we plot the best fit potential for the first and the second order Wiggly Whipped Inflation. Note that for the choice of the power $p,q$, 
the first order transition, at its best fit indicates a steep potential during the early stages of inflation, whereas the second order potential 
is relatively flat during the complete inflationary phase. The extent of the discontinuity though is not visible in the plot for the potential, 
it is evident in the slope of the potential plotted to its right. For the First order - I, we find that the derivative of the potential contains 
kink at the transition with finite width whereas First order - II represents a sharp transition in the slope. The Second order, on the 
other hand does not reflect a large change in slope but indicates a discontinuity at the transition. The middle left panel contains 
the first slow roll parameter  $\epsilon_{\rm H}=-\dot{H}/H^2$ for all the three cases for the best fits. Middle right panel 
contains the second slow roll parameter $\eta_{\rm H}=\d\ln\epsilon_{\rm H}/\d N$. $\epsilon_{\rm H},~\eta_{\rm H}$ are plotted 
as a function of {\it e-folds} from the end of inflation $N_{\rm end}$. We find that First order - I and second order nearly indicates the same 
time of transition but First order - II chooses an early transition that attempts to fit spurious features in the high-$\ell$ Planck TT data.
The second slow roll parameter clearly distinguishes the different scenarios depending on how the step in the potential and its derivatives are 
smoothed. The bottom left panel contains the absolute value of effective mass of the inflaton ($m_{\rm eff}$) as a function of field values. 
Note that during the slow roll part of the potential all the scenarios suggest the $m_{\rm eff}\sim 2\times 10^{13}{\rm GeV}$. However during the 
initial stages of inflation where we break the slow-roll moderately, the inflaton effective mass increase by an order of magnitude (First order case). During the 
transition, the rapidity of the transition or the sharpness of the step can increase the effective mass to even higher values and it can reach the GUT 
scale. The primordial scalar and tensor power spectra are plotted in the bottom right panel. First order - I provides a scalar suppression at large 
scales and at the same time provides a dip around $0.002~{\rm Mpc^{-1}}$ and a bump afterwards. However at larger wave-numbers the
oscillations soon dies and the scalar PPS converges to a power law form.  First order - II, imprints sharp oscillations that continue to 
the small scales with a decreasing amplitude. Here the dip and the bump in the PPS is not pronounced compared to First order - I 
around $0.002~{\rm Mpc^{-1}}$. The Second order generates a step in the scalar PPS. The discontinuity in the slope of the potential
leads to dip and bump around the same scales as in the case of First order - I and also contains oscillations with decreasing 
amplitude that continue in small scales like First order - II.

 \begin{figure*}[!htb]
\begin{center} 
\resizebox{420pt}{320pt}{\includegraphics{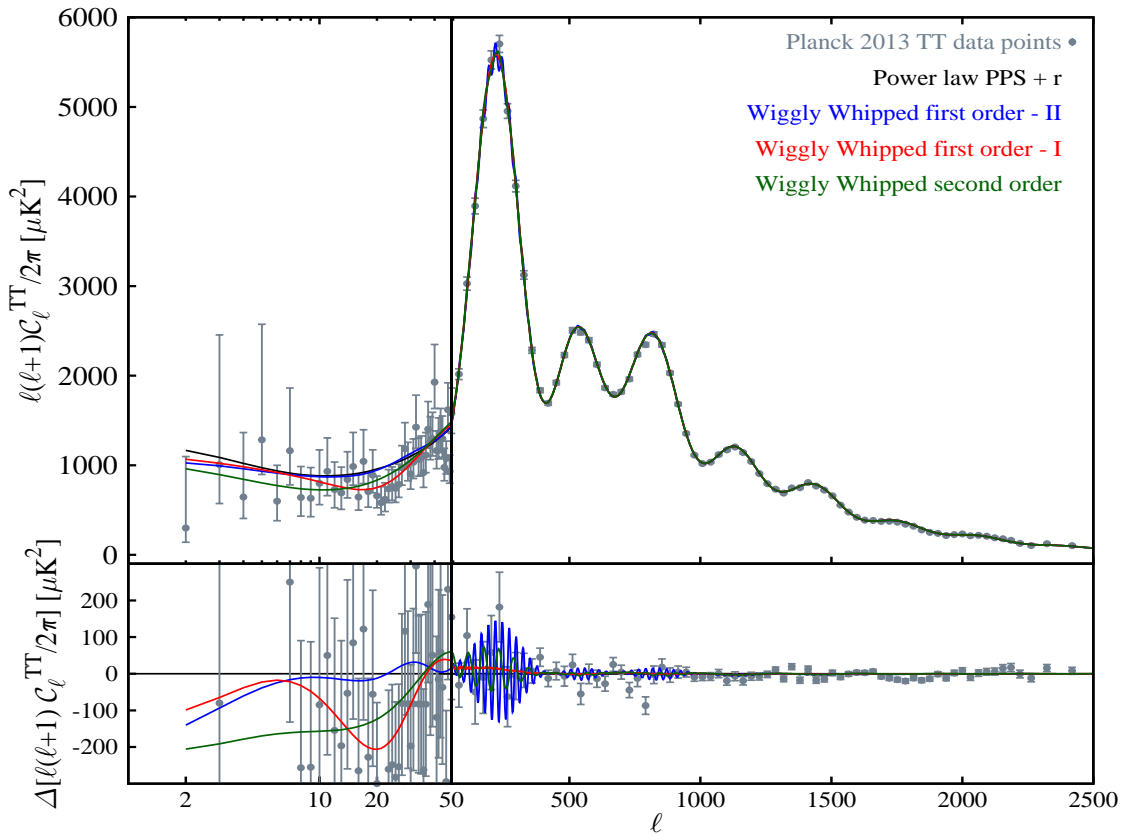}} 
\end{center}
\caption{\footnotesize\label{fig:cltt} Wiggly Whipped Inflation : Best fit $\cl^{\rm TT}$ plotted for the parameter values quoted in 
Table~\ref{tab:bestfits}. At bottom, the $\Delta\cl^{\rm TT}$, residual of the power law $\Lambda$CDM best fit model are plotted 
both for data and the Wiggly Whipped Inflation. Note that in all the cases, Wiggly Whipped Inflation is providing a large scale suppression
along with intermediate wiggles in the angular power spectra.}
\end{figure*}

In Figure.~\ref{fig:cltt} we present the angular power spectra ($\cl^{\rm TT}$) for temperature anisotropy obtained for 
the three models described above along with the $\cl^{\rm TT}$ for best fit power law PPS. The Planck data is plotted for the comparison.
Note that the Wiggly Whipped Inflation, for all the cases provide suppression in the large scales and at the same time generates wiggles
in the angular power spectra which helps to fit the Planck data better than the power law PPS. In the same plot, at the bottom we plot the 
residual angular power spectra $\Delta\cl^{\rm TT}=\cl^{\rm TT}|_{\rm Model/data}-\cl^{\rm TT}|_{\rm Power~law~\Lambda CDM}$. 
In the residual space the features in the data are clearly visible and it is evident that Wiggly Whipped Inflation model for different
transitions are sensitive to different features in the data. First order - I provides a scalar suppression at large scales and 
at the same time fits the drop and excess in power around $\ell\sim22$ and $40$ respectively.  First order - II mostly affects the 
high-$\ell$ ($\ell\ge50$) angular power spectrum and it can be seen that violent oscillations around $\ell\sim 200-250$, 500 and 750-800
addresses the features in the data around that region, which is unaddressed by the power law $\Lambda$CDM model. Second order 
Wiggly Whipped potential provides the strongest suppression in the $\cl^{\rm TT}$ at large scales and attempts to fit $\ell\sim22$ and $40$
features. Around the first CMB peak, the second order also introduces oscillations which again helps to fit the data better. However, afterwards
the amplitude of the oscillations decreases which makes the smaller scale $\cl^{\rm TT}$ very similar to the one obtained from 
power law PPS. 

\begin{figure*}[!htb]
\begin{center} 
\resizebox{210pt}{160pt}{\includegraphics{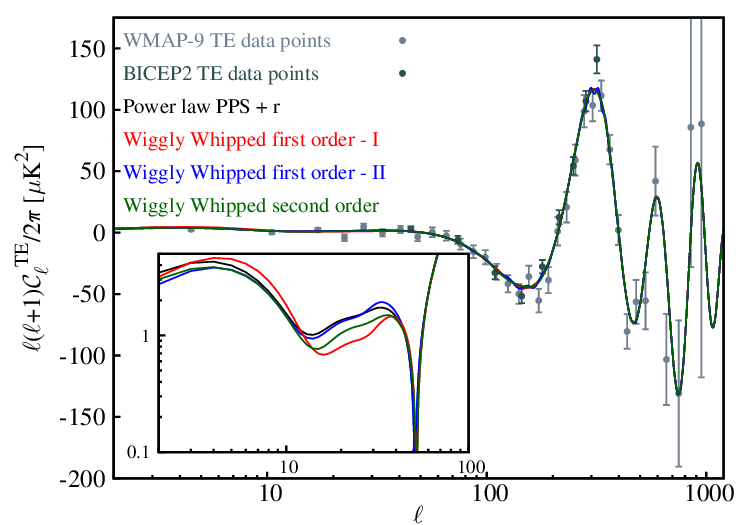}} 
\resizebox{210pt}{160pt}{\includegraphics{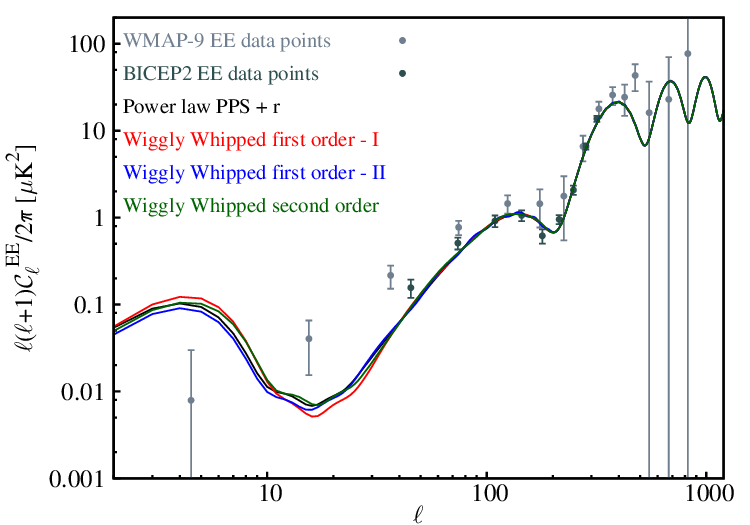}} 
\resizebox{210pt}{160pt}{\includegraphics{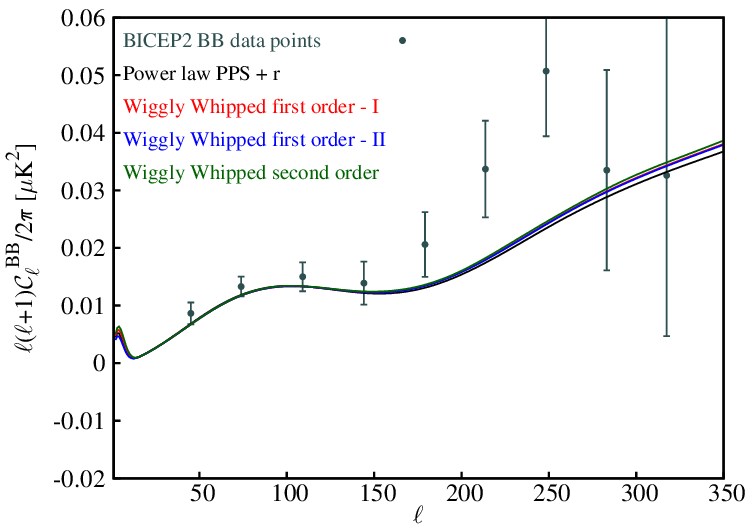}} 
\end{center}
\caption{\footnotesize\label{fig:clpol} Wiggly Whipped Inflation : TE (top left), EE (top right) and BB (bottom) polarization angular power spectra for different models and the data from WMAP-9 and BICEP2 are plotted. In the top left panel, the inset contains the absolute values of TE angular power spectra.}
\end{figure*}

In Figure~\ref{fig:clpol}, for the same models we plot the polarization 
power spectra, {\it i.e.,} $\cl^{\rm TE/EE/BB}$. The data points from WMAP 9 year
observations and BICEP2 observations are also plotted. Note that large scale suppression 
in scalar PPS are also reflected in TE/EE polarization data. Compared to WMAP-9 it is clear 
that BICEP2 data points are much closer to the model predictions in all the cases. From the plot of $\cl^{\rm BB}$, 
it is clear that all these models are able to fit the B-mode data to similar extent. We must mention here that though the 
wiggles in the First order - II address the TT data better than power law, there is a possibility that these oscillations fit 
noise in the angular power spectrum. However, similar features in the polarization spectrum will help us to distinguish real
features from the random fluctuations. We expect with Planck polarization data, we shall be able hunt down the features in the 
data with much higher confidence. 

\subsection{Non-Gaussianity}

Along with the power spectra which represents the two point correlations of perturbations, we have constraints on the bispectrum from the 
Planck bispectrum measurements. Based on Maldacena formalism~\cite{maldacena-2003} and following the methods described 
in~\cite{chen,Martin:2011sn,Hazra:2012yn,Hazra:2013nca} we use the publicly available code {\tt BINGO} to calculate the bispectrum, specifically
the local $\fnl$ in equilateral triangular configuration~\footnote{Local $\fnl$ can be derived from bispectrum 
$\cB_{_{\mathrm{S}}}(\vka,\vkb,\vkc)$ as \beq
\fnl(\vka,\vkb,\vkc)=-\frac{10}{3}\, (2\,\pi)^{-4}\; (2\,\pi)^{9/2}\;
k_{1}^3\, k_{2}^3\,k_{3}^3\; \cB_{_{\mathrm{S}}}(\vka,\vkb,\vkc)\nn\\
\times\l[k_1^{3}\; {P}_{_{\mathrm{S}}}(k_2)\; 
{P}_{_{\mathrm{S}}}(k_3)
+{\mathrm{two~permutations}}\r]^{-1}
\eeq and we use $k_1=k_2=k_3=k$ for equilateral triangular case.} for the Wiggly Whipped Inflation. 
The constraints on primordial non-Gaussianity has been found 
to be $\fnl= 2.7\pm5.8$. For Whipped Inflation we have demonstrated that $\fnl$ is ${\cal O}(0.1-0.2)$ which is certainly 
favored by Planck data.

\begin{figure*}[!htb]
\begin{center} 
\resizebox{210pt}{160pt}{\includegraphics{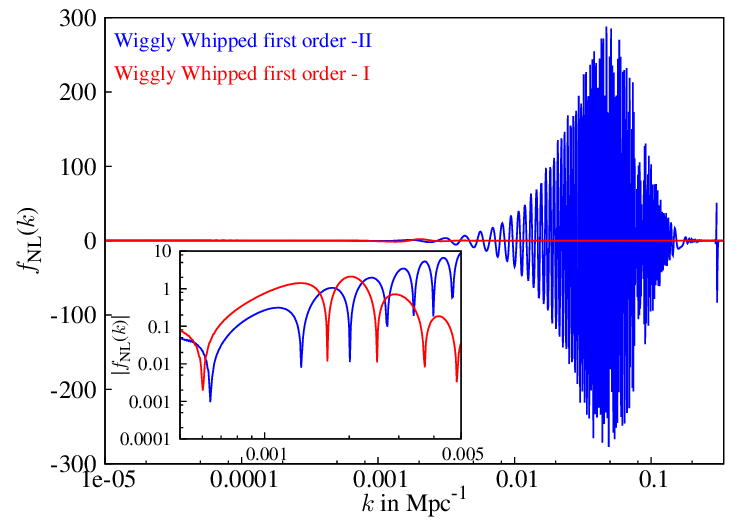}} 
\resizebox{210pt}{160pt}{\includegraphics{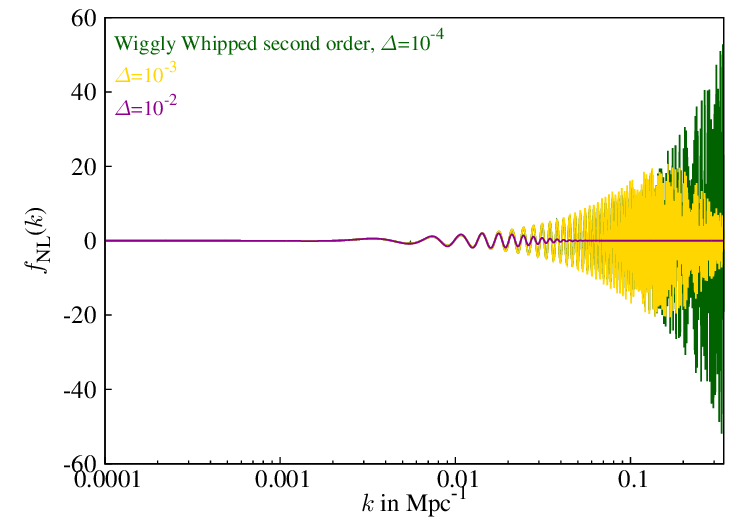}} 
\end{center}
\caption{\footnotesize\label{fig:fnlplot} Wiggly Whipped Inflation : Bi-spectrum, specifically the $\fnl$ plotted in equilateral triangular 
configuration. [Left] The $\fnl$ plotted for two first order transitions. [Right] The $\fnl$ plotted for the second order transition. For the 
second order transition, we plot the $\fnl$ for different smoothing width of the transition or equivalently for different time taken by the 
scalar field during the transition.}
\end{figure*}

However, for Wiggly Whipped Inflation one can expect higher non-Gaussianity due to sharp departures from slow roll. 
We present the $\fnl$ for the first order and second order Wiggly Whipped potentials in Figure~\ref{fig:fnlplot}. 
In left panel we plot the $\fnl$ for First order - I and II cases. In the right panel we plot the Second order case with different widths of 
smoothing. For the First order - I case, we find that the $\fnl$ is ${\cal O}(2-3)$ and for First order - II case, due to violent oscillations 
the $\fnl$ is boosted up to ${\cal O}(300)$. The inset of the left plot contains the absolute value of $\fnl$ to demonstrate the order of magnitude
difference in these two first order transitions. 

For the Second order Wiggly Whipped potential, we know that the discontinuity in the slope of the potential does not affect the primordial 
spectrum significantly but since $\fnl$ contains second derivative of the potential, which contains a Dirac Delta function, for an instantaneous 
transition this model generates a linearly divergent bispectrum for this model as has been shown before~\cite{Hazra:2012yn,Arroja:2012,Martin:2014}. 
However, as have been argued in~\cite{Arroja:2012,Martin:2014}, for any physically plausible transition, which occurs in a finite time, the bispectrum 
ceases to be divergent. This statement also holds true for the First Order -I and II, however, since the power spectrum in the first order 
transition is directly related to the smoothness of the step, we plot the bispectrum for the best fit smoothing width. In the Second order case 
we plot the $\fnl$ for three different smoothing width $\Delta$ (here $\Delta$ denotes the width in field space in ${\rm M_{Pl}}$). We note that 
for a very sharp transition, $\Delta=10^{-4}$ the $\fnl$ becomes linearly divergent, while for $\Delta=10^{-3}$ we find $\fnl$ reaches a 
maximum value of 20 and then decreases, as has also been shown in~\cite{Martin:2014}. For even smoother transition ($\Delta=10^{-2}$), we find 
$\fnl$ becomes ${\cal O}(2)$ at its peak~\footnote{For a recent discussion on bispectra generated in the models with discontinuity in 
the derivatives of the potential, see~\cite{Romano:2014kla,Martin:2014}}. We should mention that for these values of $\Delta$, the primordial 
power spectra and hence the angular 
power spectra remains unaffected. 

Now, the question is whether the large values of the primordial bispectra are supported by the Planck data. To answer this we need to compare
the angular bispectra obtained from these models with the Planck bispectrum data. From a first look, it may be argued that since 
the violent oscillations in the primordial power spectrum leads to a large and oscillating $\fnl$, in a finite bin width of Planck resolution 
the $\fnl$ will be averaged out. However, to have a full understanding of the issue, we need to wait for Planck 
polarization data to confirm the oscillations in Wiggly Whipped First order - II case and then compare the bispectrum directly. Till then 
we can at least say that Wiggly Whipped First order - I and Second order (for a smooth transition) are completely consistent with Planck bounds.


\subsection{Matter power spectra}
\begin{figure*}[!htb]
\begin{center} 
\resizebox{210pt}{160pt}{\includegraphics{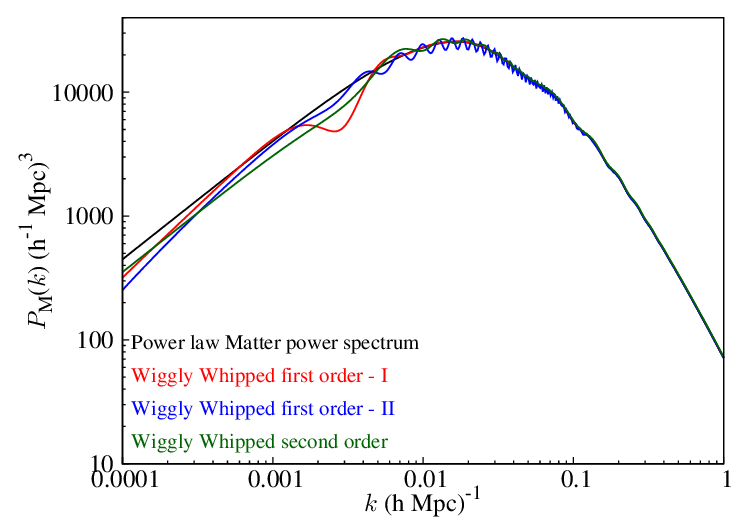}} 
\resizebox{210pt}{160pt}{\includegraphics{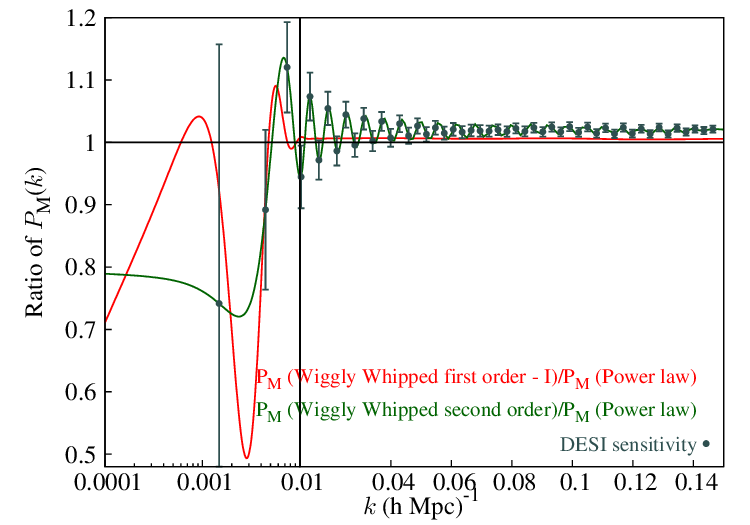}} 
\end{center}
\caption{\footnotesize\label{fig:mps} Wiggly Whipped Inflation : Matter power spectra (left) obtained from the 
best fit potential and background parameters (in Table~\ref{tab:bestfits}) and the ratio (right) {\it w.r.t.} 
the matter power spectra obtained from power law best fit model. The DESI forecasted fractional errors are 
overlayed in the right panel as well. Note that from the future matter power spectrum data we shall be able 
to identify specific features in the primordial power spectrum.}
\end{figure*}

Features in the primordial power spectra that are not located only in the large scales can alter 
the matter power spectrum in the observable range of recent and future large scale structure data. 
Wiggly Whipped Inflation introduces wiggles in the primordial power spectra that are not just located 
in the largest scales. Due to the non-local nature of the wiggles, in future it might be possible to identify 
and constrain them from matter power spectra data from DESI with high confidence. 
In Fig.~\ref{fig:mps}, we provide the matter power spectra for the Wiggly Whipped potentials 
(Eqs.~\ref{eq:potential_o1} and~\ref{eq:potential_o2}) for the best fit values of the potentials and 
the corresponding cosmological parameters provided in Table~\ref{tab:bestfits}. The left panel of the 
Fig.~\ref{fig:mps} contains the matter power spectra for different models and right panel contains the 
ratio of matter power spectra for different Wiggly Whipped models with respect to the matter power spectrum 
obtained from power law scalar PPS. Note that the First Order-I and the Second Order case imprints oscillations 
affecting a broad range from the very large scales till the baryon acoustic oscillations scales. This long 
range deviation from the expectations of the power-law PPS might be detectable by the future large scale 
structure surveys such as DESI~\cite{DESI}. 

The fractional error estimates from  for DESI~\footnote{Private communication with Pat Mc Donald.} 
are provided in the right panel. Note that  we have overlayed the errors on the Second Order transition, 
since this model provides a long range oscillations with sufficiently large width for detection. 
For the First Order - I, the large scale dip might be also well constrained with DESI. 
Independent detection of such large scale features with DESI would significantly increase 
confidence in the Wiggly Whipped scenarios as a robust model of inflation especially if CMB data
continues to fit these models well. Second Order model contains both the large scale dip and oscillations 
extending till BAO scale and it will probably be more tightly constrained since around BAO scales we shall have 
more control on the data due to low cosmic variance and large amount of cleaner data. 
Note that both First Order -I and the Second Order Wiggly Whipped is showing an excess in power at small 
scales compared to power law PPS. This is happening due to the fact that the best fit parameters are obtained 
only from CMB data, where no large scale matter power spectra data has been used.

\begin{figure*}[!htb]
\begin{center} 
\resizebox{210pt}{160pt}{\includegraphics{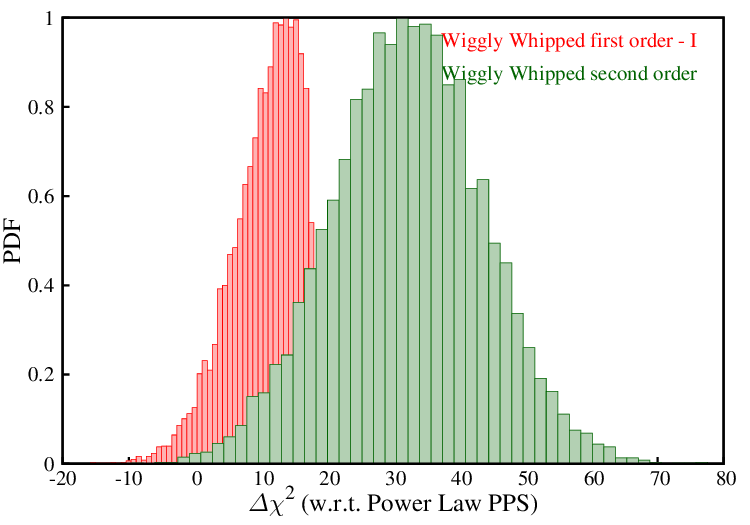}} 

\end{center}
\caption{\footnotesize\label{fig:mpkdev} 
Probability distribution function of $\Delta\chi^2=-2\l[\ln[{\cal L}_{\rm Power~law}]-\ln[{\cal L}_{\rm Wiggly~Whipped~fiducial}]\r]$ for the two cases of Wiggly Whipped first order - I and Wiggly Whipped second order are plotted. 10000 realizations of the future DESI binned matter power spectrum data have been used in these simulations. One can clearly see that the future matter power spectrum data will be sensitive to the features of the primordial spectrum providing us additional hints towards deviations from the standard power-law form of the PPS.    
}
\end{figure*}

In order to have a quantitative estimate of the expected sensitivity of DESI towards determining 
the features in the scalar PPS, we generate 10000 mock data using the projected errors in different 
scales based on two fiducial models. We have used Wiggly Whipped first order - I and second order as 
fiducial models since these two models have features which will not be averaged out in a detectable bin width. 

For the 10000 data realizations we obtain the likelihood from the corresponding fiducial model 
($\ln[{\cal L}_{\rm Wiggly~Whipped~fiducial}$) and the power law ($\ln[{\cal L}_{\rm Power~law}]$) 
allowing an overall amplitude shift. The probability distribution function (PDF) of the likelihood difference 
$\Delta\chi^2\equiv-2\Delta\ln[{\cal L}]=-2\l[\ln[{\cal L}_{\rm Power~law}]-\ln[{\cal L}_{\rm Wiggly~Whipped~fiducial}]\r]$ is plotted 
in Fig~\ref{fig:mpkdev}. One can see that the future matter power spectrum large scale structure data can indeed distinguish the Wiggly Whipped second order model and power-law with a very high confidence. It is also evident that the future matter power spectrum data can give us clear hints for the case of Wiggly Whipped first order - I or any similar case deviating from the expectations of the power-law form of the PPS.

 \subsection{Gravitational Waves}
 We have calculated the inflationary gravitational wave spectrum for our models and shown them in the bottom right panel of Figure~\ref{fig:theory}. 
 As can be seen they are tilted red relative to scale invariant spectrum, 
 which implies that they would be detectable by BBO (Big Bang Observer) but not by eLISA~\cite{elisa} or LIGO~\cite{Ligo} II or III.
 Thus, if these models continue to fit new observations, the interest in BBO gravitational wave detector becomes more strongly motivated.

\section{Conclusions}\label{sec:conclusions}
In this paper we explored a range of variations for the Whipped
inflation scenario that we had discussed in a recent
paper~\cite{Hazra:2014jka} in the light of Planck and BICEP2.
Introducing a discontinuity in the potential and/or its first
derivative at the transition point of the Whipped Inflation
potentials, we have shown that in general sharp time or field
value structures in the inflaton potential introduce Wiggles in
the primordial power spectrum as one would expect. These wiggles
can be supported by the CMB TT angular power spectrum data from
Planck. Apart from reconciling the Planck and BICEP2 data with a
large scale scalar suppression, Wiggly Whipped Inflation models go
one step further and addresses features in the TT data from
Planck. A discontinuous step introduces an instantaneous
transition in the field value or potential, which might result in
divergent two point and three point correlation function of
curvature perturbations. For a realistic transition we model the
discontinuity with a smoothed step in the inflaton potential (WWI
type I, or first order) or by a rapid change of its first derivative (WWI type
II, or second order). By comparing with Planck and BICEP2 data we show that in 
both the cases we get significant improvement in likelihood compared to
power law scenario. For two different smoothing widths we obtain
two different kinds of scalar PPS ; one of which fits the
low-$\ell$ broad features and the other attempts to fit features
in the high-$\ell$ Planck TT data.

These best-fitted large field models are found to have a
transition from a faster roll to the slow roll $V(\phi)=(1/2)m^2
\phi^2$ inflation at a field value around 14.6$M_{\rm Pl}$ and thus a
potential energy of $V(\phi) \sim (10^{16}\,{\rm
GeV})^4$. In general this transition and any features in the large
field potential produces not only suppression of scalars relative
to tensor modes at small $k$ but also introduces wiggles in the
primordial perturbation spectrum. These wiggles can help fitting
localized features in the CMB angular power spectrum and can
affect other cosmological observables too.

On the other hand, the WWI type II introduces large-scale scalar
suppression by generating a step in the large scale TT spectrum.
Along with the step this model also introduces oscillations in the
intermediate scales. This step and the oscillations helps to fit
both the large and intermediate scale temperature data from Planck
significantly better than the power law scalar PPS. The presence
and importance of such wiggles/oscillations can be tested by the
future polarization data as well. In this regard Planck
polarization data can be very insightful. We should note that for
the purpose of cosmological parameter estimation using other CMB
observations such as B-mode polarization data from
POLARBEAR~\cite{Ade:2014afa} and SPT~\cite{Hanson:2013hsb} can be
also very useful. We also calculated the extent of
non-Gaussianities, especially the bispectra in these models and we
found that for the models we considered, for a finite-time
transition, the $\fnl$ is consistent with Planck bounds. For a
very sharp transition, where the $\fnl$ is large but oscillating
rapidly around zero, the agreement with the data needs to be
checked with more carefully binned Planck bispectrum data.

Moreover, we also present the matter power spectrum in these
models at their best fit values. Using the forecasted errors from
DESI, we argue that the matter power spectrum constraints will
certainly be able to confirm the existence of the wiggly features
in the primordial power spectrum if they are really present. If
such models become highly favored by the data then one can think
of next generation of large scale structure surveys to provide
improved sensitivity to the low-$k$ region.

Here we can summarize that with the full set of CMB temperature
and E and B-mode polarization anisotropy spectra, we can determine
the inflaton potential, its slope and, most importantly for
particle physicists, the effective inflaton mass without needing
to know the underlying microscopic field (string, M-, etc.)
theory. In particular, in WWI this mass, while being almost
constant and $\sim 2\times 10^{13}$ during the last 50 {\it
e-folds} of inflation, grows and becomes of the order of $10^{14}$
GeV and higher when the inflaton field reaches the value
$\phi=\phi_0$ where its potential has a sharp feature. We have shown that 
around the sharp feature the effective mass of inflation can reach the 
GUT scale depending on the rapidity of the transition.
All this follows from cosmological observational data, leaving to
theoreticians to extract this complicated and fine mass spectrum
from particle physics at such high energies.


\section*{Acknowledgments}
D.K.H. and A.S. wish to acknowledge support from the Korea Ministry of Education, Science and Technology, Gyeongsangbuk-Do and Pohang 
City for Independent Junior Research Groups at the Asia Pacific Center for Theoretical Physics. G.F.S. acknowledges the financial support 
of the UnivEarthS Labex program at Universit\' e Sorbonne Paris Cit\' e (ANR-10-LABX-0023 and ANR-11-IDEX-0005-02). 
We thank Pat McDonald for providing the DESI matter power spectrum error estimates.
We also acknowledge the use of publicly available {\tt CAMB} and {\tt COSMOMC} in our analysis.
The authors would like to thank Antony Lewis for providing us the new {\tt COSMOMC} package that takes into account the recent BICEP2 data. 
We acknowledge the use of WMAP-9 data and from Legacy Archive for Microwave Background Data Analysis (LAMBDA)~\cite{lambdasite}, Planck
data and likelihood from Planck Legacy Archive (PLA)~\cite{PLA} and BICEP2 data from~\cite{biceprepo}. 
A.S. would like to acknowledge the support of the National Research Foundation of Korea (NRF-2013R1A1A2013795). 
A.A.S. was partially supported by the grant RFBR 14-02-00894.



\begin{thebibliography}{99}


\bibitem{BICEP2:datasets} 
  P.~A.~R.~Ade {\it et al.}  [BICEP2 Collaboration],
  arXiv:1403.4302 [astro-ph.CO].

  
\bibitem{BICEP2:Detection} 
  P.~A.~R.~Ade {\it et al.}  [BICEP2 Collaboration],
  arXiv:1403.3985 [astro-ph.CO].


\bibitem{Planck:lilelihood} 
  P.~A.~R.~Ade {\it et al.}  [Planck Collaboration],
  arXiv:1303.5075 [astro-ph.CO].
  
  \bibitem{Starobinsky:1979}
A.~Starobinsky, JETP Lett. 30, 682 (1979).  

\bibitem{Peiris:2003ff} 
  H.~V.~Peiris {\it et al.}  [WMAP Collaboration],
  Astrophys.\ J.\ Suppl.\  {\bf 148}, 213 (2003)
  [astro-ph/0302225].
  
\bibitem{Hazra:concordance} 
  D.~K.~Hazra and A.~Shafieloo,
  JCAP {\bf 01}, 043 (2014)
  [arXiv:1401.0595 [astro-ph.CO]].
 
  \bibitem{Hazra:recent}  
  D.~K.~Hazra, A.~Shafieloo, G.~F.~Smoot and A.~A.~Starobinsky,
  arXiv:1403.7786 [astro-ph.CO].

\bibitem{Hazra:2014jka} 
  D.~K.~Hazra, A.~Shafieloo, G.~F.~Smoot and A.~A.~Starobinsky,
  arXiv:1404.0360 [astro-ph.CO].
\bibitem{Levi:2013gra} 
  M.~Levi {\it et al.}  [DESI Collaboration],
  arXiv:1308.0847 [astro-ph.CO].

\bibitem{DESI}
See {\tt http://desi.lbl.gov}.
\bibitem{Contaldi:2014zua} 
  C.~R.~Contaldi, M.~Peloso and L.~Sorbo,
  arXiv:1403.4596 [astro-ph.CO].

\bibitem{Miranda:2014wga} 
  V.~íc.~Miranda, W.~Hu and P.~Adshead,
  arXiv:1403.5231 [astro-ph.CO].

\bibitem{Abazajian:2014tqa} 
  K.~N.~Abazajian, G.~Aslanyan, R.~Easther and L.~C.~Price,
  arXiv:1403.5922 [astro-ph.CO].
 
  \bibitem{Hu:2014aua} 
  B.~Hu, J.~-W.~Hu, Z.~-K.~Guo and R.~-G.~Cai,
  arXiv:1404.3690 [astro-ph.CO].




\bibitem{Kawasaki:2014fwa} 
  M.~Kawasaki, T.~Sekiguchi, T.~Takahashi and S.~Yokoyama,
  arXiv:1404.2175 [astro-ph.CO].

\bibitem{Freivogel:2014hca} 
  B.~Freivogel, M.~Kleban, M.~R.~Martinez and L.~Susskind,
  arXiv:1404.2274 [astro-ph.CO].

\bibitem{Bousso:2014jca} 
  R.~Bousso, D.~Harlow and L.~Senatore,
  arXiv:1404.2278 [astro-ph.CO].


\bibitem{Firouzjahi:2014fda} 
  H.~Firouzjahi and M.~H.~Namjoo,
  arXiv:1404.2589 [astro-ph.CO].


\bibitem{Kinney:2014jya} 
  W.~H.~Kinney and K.~Freese,
  arXiv:1404.4614 [astro-ph.CO].

\bibitem{Zibin:2014iea} 
  J.~P.~Zibin,
  arXiv:1404.4866 [astro-ph.CO].


\bibitem{Achucarro:2014msa} 
  A.~Achucarro, V.~Atal, B.~Hu, P.~Ortiz and J.~Torrado,
  arXiv:1404.7522 [astro-ph.CO].




\bibitem{Kim:2014rwa} 
  J.~E.~Kim,
  arXiv:1405.0221 [hep-th].

\bibitem{Kallosh:2014xwa} 
  R.~Kallosh, A.~Linde and A.~Westphal,
  arXiv:1405.0270 [hep-th].

\bibitem{Mukohyama:2014gba} 
  S.~Mukohyama, R.~Namba, M.~Peloso and G.~Shiu,
  arXiv:1405.0346 [astro-ph.CO].






\bibitem{Dvorkin:2014lea} 
  J.~-F.~Zhang, Y.~-H.~Li and X.~Zhang,
  arXiv:1403.7028 [astro-ph.CO];
  C.~Dvorkin, M.~Wyman, D.~H.~Rudd and W.~Hu,
  arXiv:1403.8049 [astro-ph.CO].

\bibitem{Ashoorioon:2014nta} 
  A.~Ashoorioon, K.~Dimopoulos, M.~M.~Sheikh-Jabbari and G.~Shiu,
  arXiv:1403.6099 [hep-th].
  
\bibitem{Ade:2013uln} 
  P.~A.~R.~Ade {\it et al.}  [Planck Collaboration],
  arXiv:1303.5082 [astro-ph.CO].



  \bibitem{reconstruction-all}
  S.~Hannestad,
  Phys.\ Rev.\ D {\bf 63} (2001) 043009
  [astro-ph/0009296];
  M.~Tegmark and M.~Zaldarriaga,
  Phys.\ Rev.\ D {\bf 66} (2002) 103508
  [astro-ph/0207047];
%
%
  S.~L.~Bridle, A.~M.~Lewis, J.~Weller and G.~Efstathiou,
  Mon.\ Not.\ Roy.\ Astron.\ Soc.\  {\bf 342} (2003) L72
  [astro-ph/0302306];
  A.~Shafieloo and T.~Souradeep,
  Phys.\ Rev.\ D {\bf 70} (2004) 043523
  [astro-ph/0312174];
  P.~Mukherjee and Y.~Wang,
  Astrophys.\ J.\  {\bf 599} (2003) 1
  [astro-ph/0303211];
%
%
%
%
%
  D.~Tocchini-Valentini, Y.~Hoffman and J.~Silk,
  Mon.\ Not.\ Roy.\ Astron.\ Soc.\  {\bf 367} (2006) 1095
  [astro-ph/0509478];
%
%
  N.~Kogo, M.~Sasaki and J.~'i.~Yokoyama,
  Prog.\ Theor.\ Phys.\  {\bf 114} (2005) 555
  [astro-ph/0504471];
%
%
  S.~M.~Leach,
  Mon.\ Not.\ Roy.\ Astron.\ Soc.\  {\bf 372} (2006) 646
  [astro-ph/0506390];
  A.~Shafieloo and T.~Souradeep,
  Phys.\ Rev.\ D {\bf 78} (2008) 023511
  [arXiv:0709.1944 [astro-ph]];
%
  P.~Paykari and A.~H.~Jaffe,
  Astrophys.\ J.\  {\bf 711} (2010) 1
  [arXiv:0902.4399 [astro-ph.CO]];
%
%
  G.~Nicholson and C.~R.~Contaldi,
  JCAP {\bf 0907}, 011 (2009)
  [arXiv:0903.1106 [astro-ph.CO]];
%
%
%
%
%
  C.~Gauthier and M.~Bucher,
  JCAP {\bf 1210}, 050 (2012)
  [arXiv:1209.2147 [astro-ph.CO]];
%
 R.~Hlozek, J.~Dunkley, G.~Addison, J.~W.~Appel, J.~R.~Bond, C.~S.~Carvalho, S.~Das and M.~Devlin {\it et al.},
  Astrophys.\ J.\  {\bf 749} (2012) 90
  [arXiv:1105.4887 [astro-ph.CO]];
%
  J.~A.~Vazquez, M.~Bridges, M.~P.~Hobson and A.~N.~Lasenby,
  JCAP {\bf 1206}, 006 (2012)
  [arXiv:1203.1252 [astro-ph.CO]];
    D.~K.~Hazra, A.~Shafieloo and T.~Souradeep,
  JCAP {\bf 1307}, 031 (2013)
  [arXiv:1303.4143 [astro-ph.CO]];
    D.~K.~Hazra, A.~Shafieloo and T.~Souradeep,
  Phys.\ Rev.\ D {\bf 87}, 123528 (2013)
  [arXiv:1303.5336 [astro-ph.CO]];
  P.~Hunt and S.~Sarkar,
  arXiv:1308.2317 [astro-ph.CO];
    S.~Dorn, E.~Ramirez, K.~E.~Kunze, S.~Hofmann and T.~A.~Ensslin,
  JCAP {\bf 1406}, 048 (2014)
  [arXiv:1403.5067 [astro-ph.CO]].
  

 \bibitem{Hazra:2013nca} 
  D.~K.~Hazra, A.~Shafieloo and G.~F.~Smoot,
   JCAP {\bf 1312}, 035 (2013)
  arXiv:1310.3038 [astro-ph.CO].
  
  
  

  
\bibitem{step-models} 
  J.~A.~Adams, B.~Cresswell and R.~Easther,
  Phys.\ Rev.\ D {\bf 64} (2001) 123514
  [astro-ph/0102236];
    R.~Allahverdi, K.~Enqvist, J.~Garcia-Bellido and A.~Mazumdar,
  Phys.\ Rev.\ Lett.\  {\bf 97} (2006) 191304
  [hep-ph/0605035];
    L.~Covi, J.~Hamann, A.~Melchiorri, A.~Slosar and I.~Sorbera,
  Phys.\ Rev.\ D {\bf 74} (2006) 083509
  [astro-ph/0606452];
    R.~K.~Jain, P.~Chingangbam, J.~-O.~Gong, L.~Sriramkumar and T.~Souradeep,
  JCAP {\bf 0901} (2009) 009
  [arXiv:0809.3915 [astro-ph]];
    M.~J.~Mortonson, C.~Dvorkin, H.~V.~Peiris and W.~Hu,
  Phys.\ Rev.\ D {\bf 79} (2009) 103519
  [arXiv:0903.4920 [astro-ph.CO]];
    D.~K.~Hazra, M.~Aich, R.~K.~Jain, L.~Sriramkumar and T.~Souradeep,
  JCAP {\bf 1010}, 008 (2010) 
  [arXiv:1005.2175 [astro-ph.CO]];
  V.~Miranda, W.~Hu and P.~Adshead,
  Phys.\ Rev.\ D {\bf 86}, 063529 (2012)
  [arXiv:1207.2186 [astro-ph.CO]];
    M.~Benetti,
  arXiv:1308.6406 [astro-ph.CO].
  


  
\bibitem{Planck:fnl} 
  P.~A.~R.~Ade {\it et al.}  [Planck Collaboration],
  arXiv:1303.5084 [astro-ph.CO].


 


  
  
\bibitem{S92}

A.~A.~Starobinsky, JETP Lett. {\bf 55}, 489 (1992).

    
\bibitem{Linde:1998iw} 
  A.~D.~Linde,
  Phys.\ Rev.\ D {\bf 59}, 023503 (1999)
  [hep-ph/9807493].
\bibitem{Linde:1999wv} 
  A.~D.~Linde, M.~Sasaki and T.~Tanaka,
  Phys.\ Rev.\ D {\bf 59}, 123522 (1999)
  [astro-ph/9901135].
  
    
  \bibitem{JSS08}

M.~Joy, V.~Sahni, A.~A.~Starobinsky, Phys. Rev. D {\bf 77}, 023514 (2008) [arXiv:0711.1585].

\bibitem{JSSS09}

M.~Joy, A.~Shafieloo, V.~Sahni, A.~A.~Starobinsky. JCAP {\bf 0906}, 028 (2009) [arXiv:0807.3334].  

\bibitem{Bousso:2013uia} 
  R.~Bousso, D.~Harlow and L.~Senatore,
  arXiv:1309.4060 [hep-th].
%

  



\bibitem{Takamizu:2010}
  Y.~-i.~Takamizu, S.~Mukohyama, M.~Sasaki and Y.~Tanaka,
  JCAP {\bf 1006}, 019 (2010)
  [arXiv:1004.1870 [astro-ph.CO]].
  \bibitem{Martin:2011sn} 
  J.~Martin and L.~Sriramkumar,
  JCAP {\bf 1201}, 008 (2012)
  [arXiv:1109.5838 [astro-ph.CO]].
\bibitem{Arroja:2011}  
    F.~Arroja, A.~E.~Romano and M.~Sasaki,
  Phys.\ Rev.\ D {\bf 84}, 123503 (2011)
  [arXiv:1106.5384 [astro-ph.CO]].
  
\bibitem{Hazra:2012yn} 
  D.~K.~Hazra, L.~Sriramkumar and J.~Martin,
  JCAP {\bf 1305}, 026 (2013)
  [arXiv:1201.0926 [astro-ph.CO]].
\bibitem{Arroja:2012}  
    F.~Arroja and M.~Sasaki,
  JCAP {\bf 1208}, 012 (2012)
  [arXiv:1204.6489 [astro-ph.CO]].

\bibitem{Martin:2014} 
  J.~Martin, L.~Sriramkumar and D.~K.~Hazra,
  arXiv:1404.6093 [astro-ph.CO].
  

 \bibitem{cambsite}
See, {\tt http://camb.info/.}
\bibitem{Lewis:1999bs}
  A.~Lewis, A.~Challinor and A.~Lasenby,
  Astrophys.\ J.\  {\bf 538} (2000) 473
  [astro-ph/9911177].
%
  
\bibitem{powell}
  M.~J.~D.~Powell, Cambridge NA Report NA2009/06, University of Cambridge, Cambridge (2009).
  
\bibitem{cosmomcsite}
See, {\tt http://cosmologist.info/cosmomc/.}


\bibitem{Lewis:2002ah}
  A.~Lewis and S.~Bridle,
  Phys.\ Rev.\ D {\bf 66} (2002) 103511
  [astro-ph/0205436].
    


\bibitem{Hinshaw:2012fq}
  G.~Hinshaw, D.~Larson, E.~Komatsu, D.~N.~Spergel, C.~L.~Bennett, J.~Dunkley, M.~R.~Nolta and M.~Halpern {\it et al.},
  arXiv:1212.5226 [astro-ph.CO].

\bibitem{oip}
  T.~Biswas, A.~Mazumdar and A.~Shafieloo,
  Phys.\ Rev.\ D {\bf 82}, 123517 (2010)
  [arXiv:1003.3206 [hep-th]];
  R.~Flauger, L.~McAllister, E.~Pajer, A.~Westphal and G.~Xu,
  JCAP {\bf 1006}, 009 (2010)
  [arXiv:0907.2916 [hep-th]];
  M.~Aich, D.~K.~Hazra, L.~Sriramkumar and T.~Souradeep,
  Phys.\ Rev.\ D {\bf 87}, 083526 (2013)
  [arXiv:1106.2798 [astro-ph.CO]];
  D.~K.~Hazra,
  JCAP {\bf 1303}, 003 (2013)
  [arXiv:1210.7170 [astro-ph.CO]];
  H.~Peiris, R.~Easther and R.~Flauger,
  arXiv:1303.2616 [astro-ph.CO];
  P.~D.~Meerburg and D.~N.~Spergel,
  Phys.\ Rev.\ D {\bf 89}, 063537 (2014)
  [arXiv:1308.3705 [astro-ph.CO]];
  R.~Easther and R.~Flauger,
  arXiv:1308.3736 [astro-ph.CO].
  
  \bibitem{maldacena-2003}
J.~Maldacena, JHEP\ {\bf 0305}, 013 (2003).

\bibitem{chen}
X.~Chen, Adv.\ Astron.\ {\bf 2010}, 638979 (2010);
X.~Chen, R.~Easther and E.~A.~Lim, JCAP {\bf 0706}, 023 (2007); JCAP 
{\bf 0804}, 010 (2008).

\bibitem{Romano:2014kla} 
  A.~E.~Romano and A.~G.~Cadavid,
  arXiv:1404.2985 [astro-ph.CO].
%






\bibitem{elisa}
See {\tt https://www.elisascience.org/}.
\bibitem{Ligo}
See {\tt http://www.ligo.caltech.edu/}.

\bibitem{Ade:2014afa} 
  P.~A.~R.~Ade {\it et al.}  [ The POLARBEAR Collaboration],
  arXiv:1403.2369 [astro-ph.CO].
\bibitem{Hanson:2013hsb} 
  D.~Hanson {\it et al.}  [SPTpol Collaboration],
  Phys.\ Rev.\ Lett.\  {\bf 111}, 141301 (2013)
  [arXiv:1307.5830 [astro-ph.CO]].
  \bibitem{lambdasite}  
See {\tt http://lambda.gsfc.nasa.gov/product/map/dr3/m$\_$products.cfm}

\bibitem{PLA}  
See, \\
{\tt http://www.sciops.esa.int/index.php?project=planck$\&$page=Planck$\_$Legacy$\_$Archive.}

\bibitem{biceprepo}  
See {\tt http://bicepkeck.org/\#data$\_$products.}


\end{thebibliography}
\end{document}